\documentclass[conference]{IEEEtran}
\usepackage[utf8]{inputenc}
\usepackage{tabu}
\usepackage{enumitem}
\usepackage[fleqn]{mathtools}
\usepackage{url}
\usepackage{subfig}
\usepackage{graphicx}
\usepackage{dblfloatfix}
\usepackage[hpos=0.7\paperwidth, 
            vpos=0.04\paperwidth,
            angle=90]{draftwatermark}

\title{A Datagram Extension to DNS over QUIC: Proven Resource Conservation in the Internet of Things}
\author{
  \IEEEauthorblockN{
    Darius Saif\IEEEauthorrefmark{1}, Ashraf Matrawy\IEEEauthorrefmark{2}
  }
  \IEEEauthorblockA{
    Carleton University, Department of Systems and Computer Engineering\IEEEauthorrefmark{1}, School of Information Technology\IEEEauthorrefmark{2}\\
    Email: {\{dariussaif\IEEEauthorrefmark{1}, ashrafmatrawy\IEEEauthorrefmark{2}}\}@cunet.carleton.ca 
  }
}
\begin{document}
\maketitle

\SetWatermarkText{Authors' Draft for Soliciting Feedback: \today}
\SetWatermarkColor[gray]{0.5}
\SetWatermarkFontSize{0.6cm}
\SetWatermarkAngle{0}
\SetWatermarkHorCenter{11cm}

\maketitle

\begin{abstract}
In this paper, we investigate the Domain Name System (DNS) over QUIC (DoQ) and propose a non-disruptive extension, which can greatly reduce DoQ's resource consumption. This extension can benefit \textit{all} DNS clients – especially Internet of Things (IoT) devices. This is important because even resource-constrained IoT devices can generate dozens of DNS requests every hour. DNS is a crucial service that correlates IP addresses and domain names. It is traditionally sent as plain-text, favoring low-latency results over security and privacy. The repercussion of this can be eavesdropping and information leakage about IoT devices. To address these concerns, the newest and most promising solution is DoQ. QUIC offers features similar to TCP and TLS while also supporting early data delivery and stream multiplexing. DoQ’s specification requires that DNS exchanges occur over independent streams in a long-lived QUIC connection. Our hypothesis is that due to DNS's typically high transaction volume, managing QUIC streams may be overly resource intensive for IoT devices. Therefore, we have designed and implemented a data delivery mode for DoQ using QUIC datagrams, which we believe to be more preferable than stream-based delivery. To test our theory, we analyzed the memory, CPU, signaling, power, and time of each DoQ delivery mode in a setup generating real queries and network traffic. Our novel datagram-based delivery mode proved to be decisively more resource-friendly with little compromise in terms of functionality or performance. Furthermore, our paper is the first to investigate multiple queries over DoQ, to our knowledge.
\end{abstract}

\IEEEpeerreviewmaketitle

\section{Introduction}

\IEEEPARstart{T}{he} Domain Name System (DNS) protocol has been a crucial service since the early days of the Internet, allowing for decentralized and scalable growth \cite{dunlapDns}. In its traditional sense, clients query a DNS server with a text-based domain name, and the server consults its hierarchical database to respond with the corresponding IP address or vice versa. This results in user-friendly resource retrieval without exposure to the underlying mechanics of the Internet. More recently, DNS has been integral in load-balancing content requests across edge servers, forming the basis of Content Delivery Networks \cite{akamaiCdn}. To put into perspective the importance of DNS in the modern Internet, Google's public DNS resolver handled more than a trillion DNS queries per day in 2018\footnote{https://security.googleblog.com/2018/08/google-public-dns-turns-8888-years-old.html}.

By nature of its original inception for low-latency results, DNS exchanges are historically communicated over plain-text. However, side effects of this can include eavesdropping, source address spoofing, and injection attacks \cite{dnsPrivacy}. DNS Security Extensions (DNSSEC) provides response data integrity and origin authenticity \cite{yangDnsSec}, but not privacy between a client and a server. Several architectures seek to address the balancing act of efficient and privacy-focused DNS \cite{rfc7858,rfc8094,rfc8484} -- the most recent is DNS over QUIC (DoQ) \cite{rfc9250}.

QUIC (not an acronym) \cite{rfc9000} is a general-purpose transport protocol that offers secure and reliable data transmission. It integrates TLSv1.3 \cite{rfc8446} to facilitate i.) early data transmission without incurring any Round-Trip Time (0-RTT) and ii.) header and data encryption. QUIC packets consist of different frame types, and packets belonging to the same connection can be sent in a single UDP datagram. One of QUIC's most appealing features is stream-multiplexing: within a single QUIC connection, multiple independent byte-streams of data can be created, unlike with TCP. Hence, the inherent issue of Head-of-Line Blocking (HoLB) with TCP can be greatly reduced by QUIC, since loss of data on one stream does not affect other streams. QUIC has already seen widespread adoption on the Internet, being HTTP/3’s \cite{rfc9114} transport.

\textbf{Our motivation} is to investigate DoQ specifically from the lens of another massive paradigm shift in the Internet -- the Internet of Things (IoT). Our hypothesis is that, under its current operation principle, DoQ may create QUIC streams in high volumes, which will be resource and signaling intensive. This is especially impactful on IoT devices due to their constrained nature. Investigations have shown not only that smart devices like lightbulbs can account for dozens of DNS queries (and thus, potential QUIC streams) every hour \cite{kuaiDnsInIot}, but also that listeners can leverage plain-text DNS messages to determine information about such devices \cite{9230403}.

\textbf{Our proposal} is a data delivery solution for DoQ using QUIC \texttt{DATAGRAM} frames, which we believe will be more resource-friendly and compromise little in terms of functionality and performance. Moreover, our proposed architecture is non-disruptive to the existing specification -- we outline a fallback mechanism where endpoints can still engage in stream-based delivery. To test our theory, we have modified an open-source DoQ implementation to run in two modes of operation: i.) stream-based, as per DoQ's specification, and ii.) our proposed datagram-based architecture.

\textbf{Our evaluation} consisted of real traffic from both transmission modes of DoQ under different network conditions and DNS transaction scenarios. Key metrics were collected from a resource-constrained device -- the Raspberry Pi Zero. The data showed that for every stream-based DNS exchange, approximately 2.5kB more memory was allocated than datagram-based. Stream-based delivery transmitted roughly 130 more bytes per DNS exchange, since \texttt{MAX\_STREAMS} frames had to be sent in accordance with QUIC's stream concurrency rules. The additional memory allocations, bytes exchanged, and stream management also led to higher CPU and power consumption. The savings of datagram-based delivery came with no statistically significant impact on performance.


\noindent \textbf{Our contributions} include:
\begin{itemize}
    \item A complete architecture (and its implementation) of a novel datagram-based delivery mode for DoQ which supplements RFC 9250 and includes a fallback mechanism to the traditional stream-based delivery mode
    \item A 5-dimensional statistical analysis of DoQ with a focus on IoT clients with limited resources
    \item Among the first research papers to investigate DoQ, and to our knowledge, the first to study multiple queries inside a QUIC connection
\end{itemize}

The rest of this paper is organized as follows: Section II covers related work motivating the integration of DoQ in IoT. Section III highlights the design features of DoQ and the benefits it offers over its contemporaries. In Section IV, we outline our proposed extension to DoQ. Sections V and VI specify our experimental methodology and results, respectively. Lastly, our conclusions are drawn in Section VII.

\section{Related Work}

Perdisci \textit{et al.}'s IoTFinder \cite{9230403} helps exemplify how plain-text DNS queries can lead to the classification of IoT devices (even if behind a NAT) with granularity potentially down to their device models. Techniques like this motivate why privacy-focused transport for DNS is important in IoT. Perdisci \textit{et al.} laid out a machine-learning framework that effectively and automatically learned the fingerprints of 50 different smart devices and filtered them out from non-IoT device traffic as well. This was done solely by inspecting DNS domain queries and their frequency. They collected data from their own lab, in addition to third-party datasets and data from an American Internet service provider.

Several encrypted DNS efforts have been proposed over the years. The motivation for such efforts is privacy: attempting to eliminate eavesdropping and the possibility of on-path tampering with DNS exchanges. Examples of this include DNS over TLS (DoT), DNS over DTLS (DoD), as well as DNS over HTTPS (DoH). These are specified under RFCs 7858 \cite{rfc7858}, 8094 \cite{rfc8094}, and 8484 \cite{rfc8484}, respectively. DoH maps DNS messages to HTTP transactions, supporting queries in both \texttt{POST} and \texttt{GET} messages. Although DoT and DoH offer privacy, this comes at the cost of higher latency and HoLB due to their underlying TCP transport. DoD, on the other hand, cannot guarantee message privacy when the DTLS record size is larger than the path Maximum Transmission Unit (MTU) \cite{rfc8094} and is limited to stub-to-resolver transactions only \cite{rfc9250}. To address these drawbacks, DoQ has been proposed.

Kosek \textit{et al.} have studied the performance of single queries via DoQ compared to DoT, DoH, and DNS over UDP for web browsers \cite{kosekDoq}. From six different geographical vantage points (one per continent), they sent queries to various public DNS resolvers and measured: i.) DNS payload size, ii.) DNS response time, and iii.) web performance for various popular websites. Kosek \textit{et al.} note in their work that 0-RTT early data support was not available on public DNS resolvers that they discovered and tested. Ultimately, they found that DoQ's single query response time was approximately 33\% faster than DoT and DoH. This equated to up to 10\% faster page loads than DoH. Lastly, they found that DoQ was only about 2\% slower than classic DNS over UDP when loading more complicated web pages. Resource utilization and multiple queries, however, were not in the scope of their work.

Batenburg \cite{batenburg2022performance} analyzed the latency between DNS over UDP, DoT, DoH, and DoQ. Delays were measured between a recursive resolver and an authoritative name server, with the \textit{dnsproxy} software package controlling the transport protocol of each query. All four candidate protocols were tested in parallel loops, each generating Record Type A DNS queries every two seconds for a period of one hour. Because these loops were run with the Linux \texttt{watch} command, DoT, DoH, and DoQ would require a new connection per query. Batenburg noted that connection establishment with QUIC's 0-RTT feature was not used either. This experimental setup was tested in a controlled environment and also hosted in the cloud to cover different geographic locations and latencies (5, 85, and 248ms). For short to medium RTTs, DoQ fared very similarly to DNS over UDP, with DoT and DoH trailing behind. For larger RTTs, DoH was found to perform slightly better than DoQ. Neither Kosek nor Batenburg's work consider low-powered devices like those present in IoT.

Lenders \textit{et al.} \cite{lenders2023securing} proposed DNS over Constrained Application Protocol (CoAP) \cite{rfc7252} for IoT. Lenders motivates this idea by stating that DoH, DoT, and DoQ (all using TLS) are quite heavy for the lowest end of constrained devices. While true, mid-tier classes of IoT devices have been proven capable of running QUIC \cite{eggert2020towards}. Since CoAP implements methods similar to HTTP, Lenders' implementation has some parallels with DoH. Lenders used the \texttt{FETCH} method to transmit DNS queries to support caching and block-wise transfers. Memory consumption, packet sizes, and resolution time were considered in their evaluation which took place for 50 DNS queries of Record Type A and AAAA at a rate of five queries per second. Their testing took place on the FIT IoT-LAB \cite{adjih2015fit}, making use of resource constrained micro-controller devices. They compared DTLS and OSCORE \cite{rfc8613} for message security: DTLS had perfect forward secrecy, but came at the cost of higher memory footprint and packet sizes. Their results for resolution time closely followed each respective method's trends regarding packet sizes. The prospect of widespread support of DNS over CoAP on the Internet remains unknown.

Our previous work \cite{saifIoTJourn} has studied the tuning of various transport parameters and APIs of the QUIC-GO library\footnote{https://github.com/quic-go/quic-go} for practical use in IoT. We generated traffic using an open-source MQTT over QUIC implementation \cite{fernandez2020and} as well as our own HTTP/3 based publish-subscribe implementation \cite{cscnpaper}. Network emulation was used to mimic the channel conditions of Narrow-Band-IoT (NB-IoT) \cite{tmobile} with Raspberry Pi Zero clients. Using eight different metrics, we provided detailed discussions on: i.) the use of X.509 certificates, ii.) choice of various timers and \textit{quic\_transport\_parameters} negotiated on connection establishment, and iii.) a high watermark scheme for endpoints sending QUIC \texttt{MAX\_STREAMS} frames. We showed that tuning of these various points had positive effects on performance, signaling overhead, and resource utilization.

We also proposed the concept of QUIC 'thin-apps', where QUIC's features were fully integrated into an application's design rather than mapping it directly to QUIC \cite{Saif_2024}. Our reasoning was that QUIC on its own is such a robust transport that many application-layer complexities may be reduced or even eliminated. We designed a publish-subscribe model which used uni-directional QUIC streams for publishing (client-initiated) and subscribing (broker-initiated) to topics. Where MQTT needed application-layer control messaging for such actions, our implementation relied solely on QUIC. Since we modeled each stream as a unique topic, topic names only needed to be sent once, resulting in lower signaling overhead compared to MQTT version 5 \cite{mqttv5}, even with topic aliasing enabled. DoQ's use-case is more of a concern than mapping topics to unique streams for publish-subscribe though, because DNS exchanges would exhaust many more QUIC streams.

\section{DNS over QUIC (DoQ)}

\begin{figure}[b]
\centering
\includegraphics[width=3.49in]{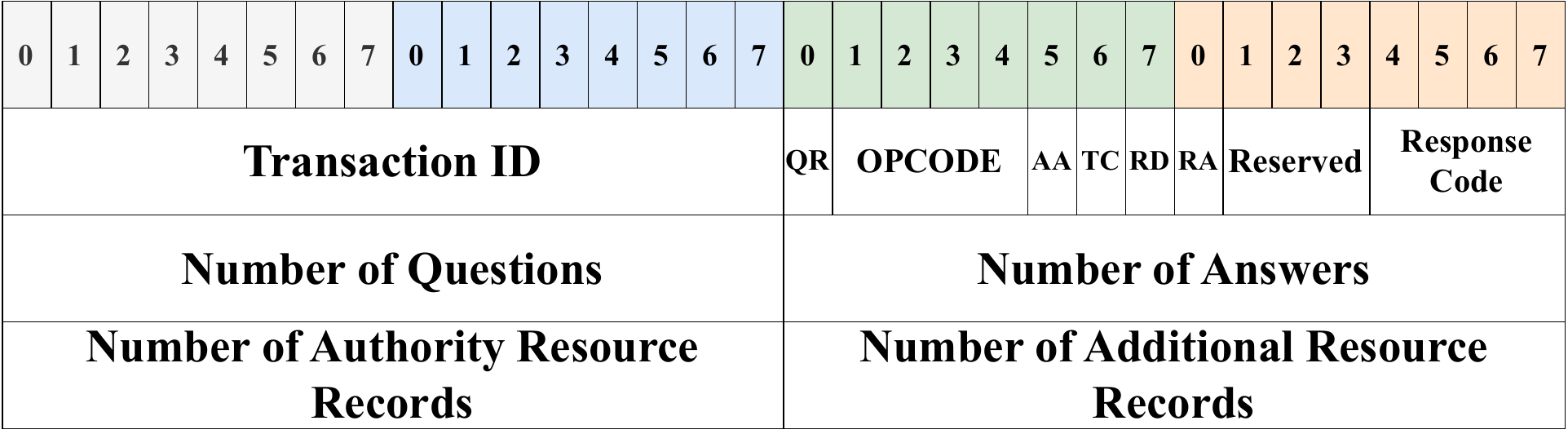}
\vspace{-4mm}
\caption{DNS Message Header}
\label{dnsHeader}
\end{figure}

DoQ is a proposed standard under RFC 9250 \cite{rfc9250}. The design principle is to map the DNS protocol, whose header is shown in Figure \ref{dnsHeader}, to QUIC transport. According to the RFC, QUIC can provide: i.) improved client and server authentication, ii.) improved source address validation for DNS servers, and iii.) fewer MTU limitations for DNS responses. Although DoT and DoH offer some of these benefits as well, QUIC's latency is closer to that of classic DNS over UDP because of its optional 0-RTT connection establishment.

RFC 9250 notes that even though DoH can make use of HTTP/3 \cite{rfc9114} (whose transport is QUIC), DoQ is a more lightweight mapping for DNS. For transfer scenarios between zones or from recursive to authoritative servers especially, HTTP's overhead proves to be too high \cite{rfc9250}. 

Several privacy considerations are also outlined in the RFC. When sending 0-RTT early data, 'replayable' DNS Operation Codes (OPCODES) must not be transmitted. Examples of replayable OPCODES are \texttt{QUERY} and \texttt{NOTIFY}. Additionally, padding must be used to obfuscate the records communicated over DNS from traffic analysis methods. This can be done either with QUIC's \texttt{PADDING} frames or with the DNS padding extension defined in RFC 7830 \cite{rfc7830}. 

DoQ proposes that long-lived QUIC connections be established between a client and DNS server and that each endpoint negotiates QUIC's idle timeout feature. Unless otherwise agreed upon by the endpoints, UDP port 853 must be used for DoQ. The advantage of using other ports (for example, 443) is that they are less likely to be blocked. Furthermore, all messages must be encoded as a 2-octet length field followed by the message content -- just like DNS over TCP. Therefore, the maximum size of a DNS message is 65535 bytes.

For each unique DNS transaction, separate QUIC streams must be used. Particularly, clients must initiate bi-directional QUIC streams for each query, and the server must respond on that same stream. This enables servers to respond to queries out of order, which helps to minimize HoLB. Once each peer has no further data to send over a QUIC stream, they must both set the stream's \texttt{FIN} bit to close the stream. If a client wishes to cancel a transaction, they can send a \texttt{STOP\_SENDING} frame for the appropriate stream. Several error codes are also defined for streams or connections that are abruptly closed.

RFC 9250 notes that because DoQ exchanges are correlated by distinct streams, the DNS Transaction ID field is technically unnecessary. Normally, the Transaction ID is a random value chosen by the initiator and is read by the responding endpoint; the same value is encoded into its response in order to correlate the query and response as a pair. As a convention, every query's Transaction ID in DoQ must be set to zero. In the case that a DoQ message be forwarded over a transport other than QUIC however, a Transaction ID must be generated.

\section{Proposed Architecture}

\subsection{Problem Framing \& Hypothesis}

Our main research objective is to determine whether DoQ, as specified by its RFC, can run reasonably well on IoT devices. Such devices tend to be limited in computational power and operate over lossy channels with unpredictable bandwidth \cite{rfc7228}. In our previous work \cite{saifIoTJourn}, which sought to tune QUIC for IoT devices, we made observations which we believe warrant consideration in DoQ's standardization:

Firstly, we found that the creation and management of QUIC streams stood out as a memory-intensive task. When 100 streams were created over the lifetime of a QUIC connection, code profiling revealed that considerably more memory was consumed compared to if only a single stream had been created \cite{saifIoTJourn}. This is a concern because the frequency of DNS queries, even from IoT devices, can be in the dozens every hour \cite{kuaiDnsInIot}.

Secondly, we observed that QUIC \texttt{MAX\_STREAMS} frame transmissions can affect signaling overhead and efficiency over time. \texttt{MAX\_STREAMS} frames inform an endpoint about the cumulative number of streams (bi-directional or uni-directional) that it is permitted to open. Section 4.6 of QUIC's specification \cite{rfc9000} "\textit{leaves implementations to decide when and how many streams should be advertised to a peer via \texttt{MAX\_STREAMS}. Implementations may choose to increase limits as streams are closed, to keep the number of streams available to peers roughly consistent.}" Indeed, we found that the QUIC-GO library transmits \texttt{MAX\_STREAMS} frames after every stream was retired. While these frames are not colossal in size, having more packets to send and process can further chew into precious resources like bandwidth and device battery.

For these reasons, we hypothesize that DoQ may prove quite taxing on IoT devices, both in terms of resources as well as signaling. We believe that incorporating \texttt{DATAGRAM} frames into the DoQ standard will address these deficiencies for \textit{all} DoQ clients, and the benefits will be most apparent on IoT devices. RFC 9221 \cite{rfc9221} introduces optional \texttt{DATAGRAM} type frames for QUIC packets, which can be used instead of streams to transmit application data. In their current states, neither RFC 9221 nor RFC 9250 make mention of one another.

\subsection{Theory of our Proposed Solution}

Any endpoint wishing to receive datagrams must set the \textit{max\_datagram\_frame\_size} QUIC transport parameter to a non-zero value when establishing a QUIC connection. RFC 9221 recommends that this transport parameter be set to 65535 bytes if datagram support is to be enabled. This ensures that any datagram which fits in a QUIC packet will be accepted by the endpoint. DoQ (and DNS over TCP) already imposes a 2-octet length field for all messages, meaning that implementing datagrams would not pose any new restrictions on the theoretical maximum DNS message size.

Because \texttt{DATAGRAM} frames are an optional feature of QUIC, we propose our solution as a non-disruptive approach to supplement the existing DoQ standard. If both the client and the server advertise datagram support, then our datagram-based DoQ is enabled. If either of the endpoints does not advertise datagram support, then the endpoints will fall back to QUIC streams to exchange DNS messages. That is, clients who do not wish to participate in datagram exchanges are not obliged to. We also note that it is possible for a single QUIC connection to make use of both streams and datagrams simultaneously, but we do not envision a case where this would be necessary in the context of DoQ.

All security and privacy properties of a QUIC connection hold true when datagrams are used. In particular, TLSv1.3, 0-RTT early data eligibility, padding, encrypted headers and data, as well as authentication and source address validation all remain intact. Therefore, our proposed solution can allow IoT devices to take full advantage of QUIC's rich feature-set for private DNS without the need for managing streams. 

Since the DNS header already contains the Transaction ID field, query-response pairs sent over datagrams can easily be correlated. We propose that when datagrams are used instead of dedicated streams, DoQ should assign and encode a non-zero Transaction ID, as per the DNS norm. If a DoQ message needed to be sent over another transport, no further action would be required.

Reliable delivery of \texttt{DATAGRAM} frames is not guaranteed. They are, however, \texttt{ACK} eliciting -- indicating the peer received and processed the data, but not necessarily its application layer. Section 5.2 of RFC 9221 states: \textit{"If a sender detects that a packet containing a specific \texttt{DATAGRAM} frame might have been lost, the implementation MAY notify the application that it believes the datagram was lost."} In such a case, DoQ can retransmit the query. \texttt{DATAGRAM} frames are not byte-streams and can arrive in any order, meaning HoLB is not a factor like it would be with DoT, DoH, and DNS over TCP.

Datagrams are subject to QUIC's congestion control, but not flow control. If an endpoint bombards another with datagrams, there are two options: i.) an endpoint can close the QUIC connection with error code \texttt{DOQ\_EXCESSIVE\_LOAD} under suspicion of malicious intent, or ii.) the endpoint can simply disregard the datagrams, as their reliable delivery is not guaranteed. \texttt{MAX\_STREAMS} frames help police the number of streams an endpoint is able to create, offering another layer of protection that datagram-based delivery does not. 

\subsection{Implementing Datagram-based DoQ}

We use the \textit{dnslookup}\footnote{https://github.com/ameshkov/dnslookup} and \textit{dnsproxy}\footnote{https://github.com/AdguardTeam/dnsproxy} packages towards implementing our proposed architecture and evaluating it against the DoQ specification. Both of these packages are open-source and are written in the GO language. We chose these packages because they already support DoQ and because their underlying library for QUIC support, QUIC-GO, implements \texttt{DATAGRAM} frames specified by RFC 9221.

We first introduced the QUIC-GO library's \textit{EnableDatagrams} option to the \textit{quic.Config} in the client's \textit{doq.go} and the server's \textit{quic\_server.go} files. This is a Boolean which drives the advertisement of datagram support in the \textit{quic\_transport\_parameters}. When the QUIC connection is established, we call the \textit{ConnectionState().SupportsDatagrams} API to determine whether datagrams should be used or if the endpoints should fall back to QUIC streams. 

The \textit{dnslookup} package's APIs generate a non-zero Transaction ID, and subsequently forces the value to zero in its \textit{Exchange} function in order to adhere to DoQ. We have added conditional logic in this function so that the Transaction ID remains intact if datagram-based DoQ is supported by both endpoints. If not, the original code is run. Because datagram-based delivery relies on the Transaction ID field, unlike stream-based, we also had to write code to associate received responses to their corresponding queries. 

Lastly, we applied conditional logic to \textit{doq.go}'s \textit{exchangeQUIC} and \textit{server\_quic.go}'s \textit{handleQUICConnection} and \textit{respondQUIC} functions to support both stream-based and our proposed datagram-based DoQ. Specifically, we make calls to QUIC-GO's \textit{SendDatagram} and \textit{ReceiveDatagram} functions. We note that the QUIC-GO library does not currently inform an application whether a datagram is lost or acknowledged -- this is a future work item got QUIC-GO covered by issue 4273\footnote{https://github.com/quic-go/quic-go/issues/4273}. We instead implemented a simple exponential back-off timer to retransmit queries if no response is received in time. The base delay was 200ms with a back-off factor of 2 and a maximum of five retries -- no jitter was included.

\section{Evaluation Methodology \& Setup}
Because our contributions involve extending DoQ and previous research \cite{kosekDoq, batenburg2022performance} has already compared DoQ against its contemporaries, we focus only on DoQ. In our setup, the \textit{dnslookup} package (Release v1.11.1) was responsible for initiating the QUIC connection and generating DNS queries. Because we wish to model the client as a resource-constrained IoT device, this program ran on a Raspberry Pi Zero -- a low-cost board with 512MB of RAM and a single core 1GHz ARM CPU running Raspbian OS 11. 

\begin{figure}[t]
\centering
\includegraphics[width=2.6in]{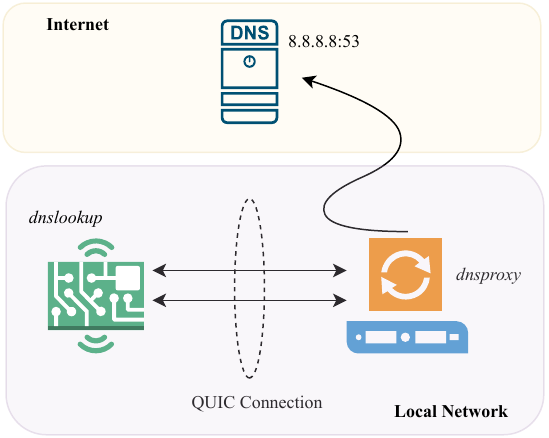}
\caption{Experimental Network Setup}
\label{netSet}
\vspace{-3mm}
\end{figure}

The \textit{dnsproxy} package (Release v0.74.1) received queries over the client's QUIC connection and referred to Google's public DNS resolver (8.8.8.8) using DNS over UDP to fulfill \textit{every} query. Then, the response was forwarded to the client over the QUIC connection. \textit{dnsproxy} ran on a Linux box and communication between the client and proxy server took place over a WiFi router. Between the client and proxy was an average delay of 94ms with a standard deviation of 64ms. This setup is illustrated in Figure \ref{netSet}.

We built and evaluated the client and server programs in two modes of operation: i.) using stream-based DoQ as specified in RFC 9250, and ii.) using our proposed datagram solution. On the Linux server, we collected packet captures using \textit{tshark} and modified the \textit{dnsproxy} package to dump the SSL keys of each QUIC connection so that the QUIC packets could be decrypted for our analysis.

The \textit{dnslookup} package only supported sending a single DNS query for a domain. To achieve multiple queries within a single QUIC connection, we modified the code to query a single domain in a loop where each query was spaced apart by 500 milliseconds. Every time the loop ran, a new Transaction ID was created for datagram-based delivery.



We also added calls to the GO language's \textit{pprof} package for memory and CPU tracing. We collected heap memory according to the program's \textit{alloc\_space} -- this measures the total number of bytes allocated since the beginning of the program, including those which were garbage-collected. The \textit{MemProfileRate} was set to 1 so that every allocated block was included in the profile. CPU traces showed how much processing time a particular function required -- the CPU profiling rate was set to 250 Hz. Traces with \textit{pprof} were collected exclusively from the client device.

To determine the resource footprint of running stream versus datagram-based DoQ, we have collected results where either 50 or 100 DNS queries were sent by the client. Given Kuai \textit{et al.}'s work studying the typical frequency of DNS queries from smart devices \cite{kuaiDnsInIot}, this represents an approximate range of what one hour's worth of DNS traffic from a resource-constrained IoT device would be. Furthermore, doubling the number of queries helps distinguish any data trends as the volume of queries increases over a QUIC connection.

We generate Address Record (A) Type DNS queries for \texttt{example.org}. In our testing, the corresponding response to each query always contained 4 answers, each consisting of a different IP address. The metrics collected were client CPU, memory, and power, as well as bytes transmitted and total transaction time. Each experiment was run for 10 iterations to ensure statistically reliable results.

Box and whisker plots were generated to illustrate various statistical features of the data. Whiskers indicate the minimum and maximum values of the inter-quartile range. The top and bottom ends of each box represent the median values of the 3rd and 1st quartiles, respectively. The line splitting the box indicates the median. Mean values are designated with an 'X' symbol and outliers with a dot.


\section{Results}

\subsection{Client Memory Usage}

\begin{figure}[b]
\centering
\vspace{-4mm}
\includegraphics[width=3.3in]{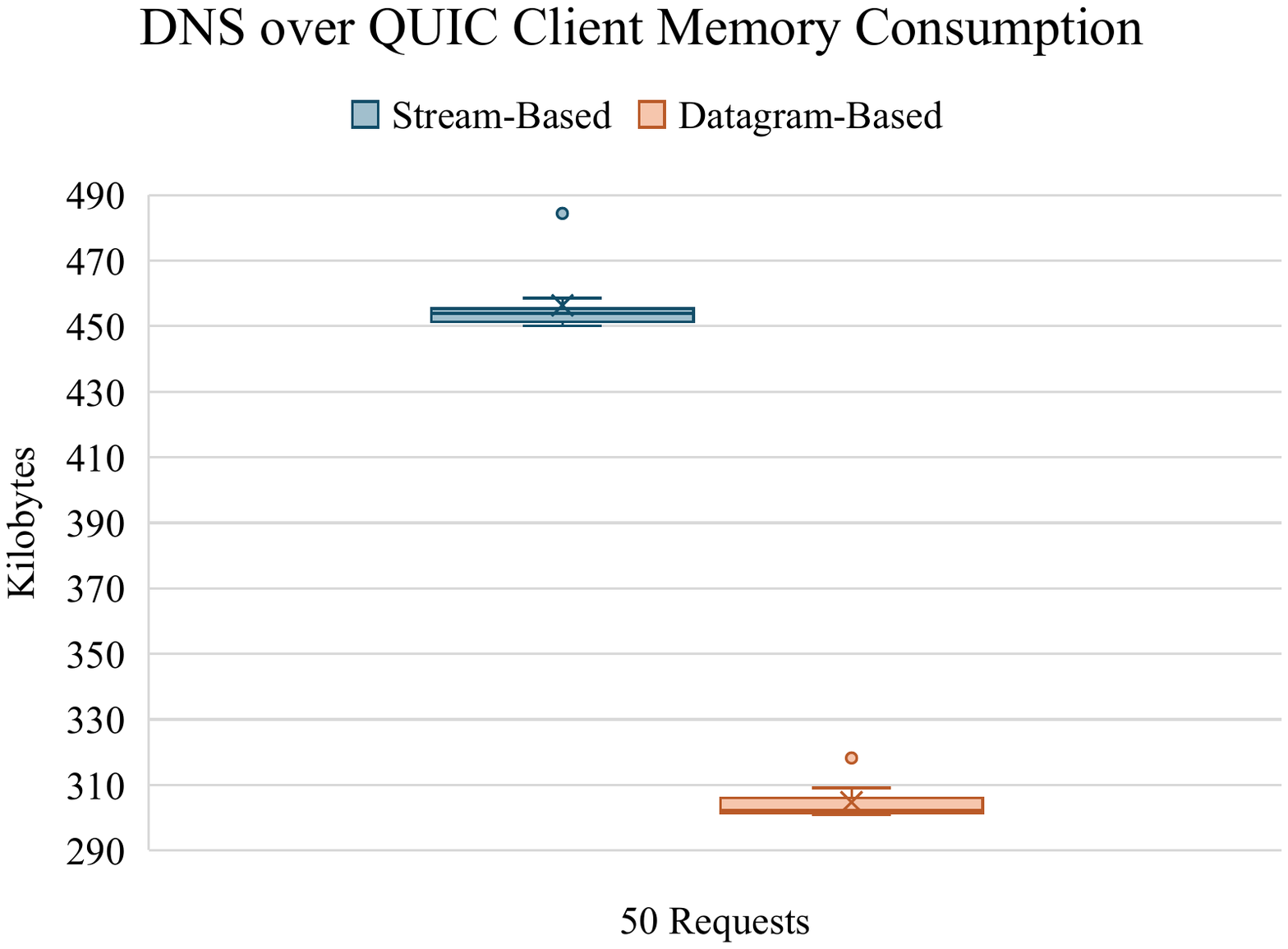}
\vspace{-2mm}
\caption{DoQ Memory Comparison (50 Requests)}
\label{mem50}
\end{figure}

The statistical features of the total allocated memory for 50 query-response pairs in a single QUIC connection are shown in Figure \ref{mem50}. Datagram-based DoQ had a decisively lower memory footprint. The memory consumption for dialing the server's address, TLS handshaking, and listening for packets was comparable between stream-based and datagram-based delivery. The entirety of the difference in memory usage was due to the \textit{exchangeQUIC} function on the client.

\begin{figure*}[t]
\centering
\vspace{-3mm}
\includegraphics[width=7.14in]{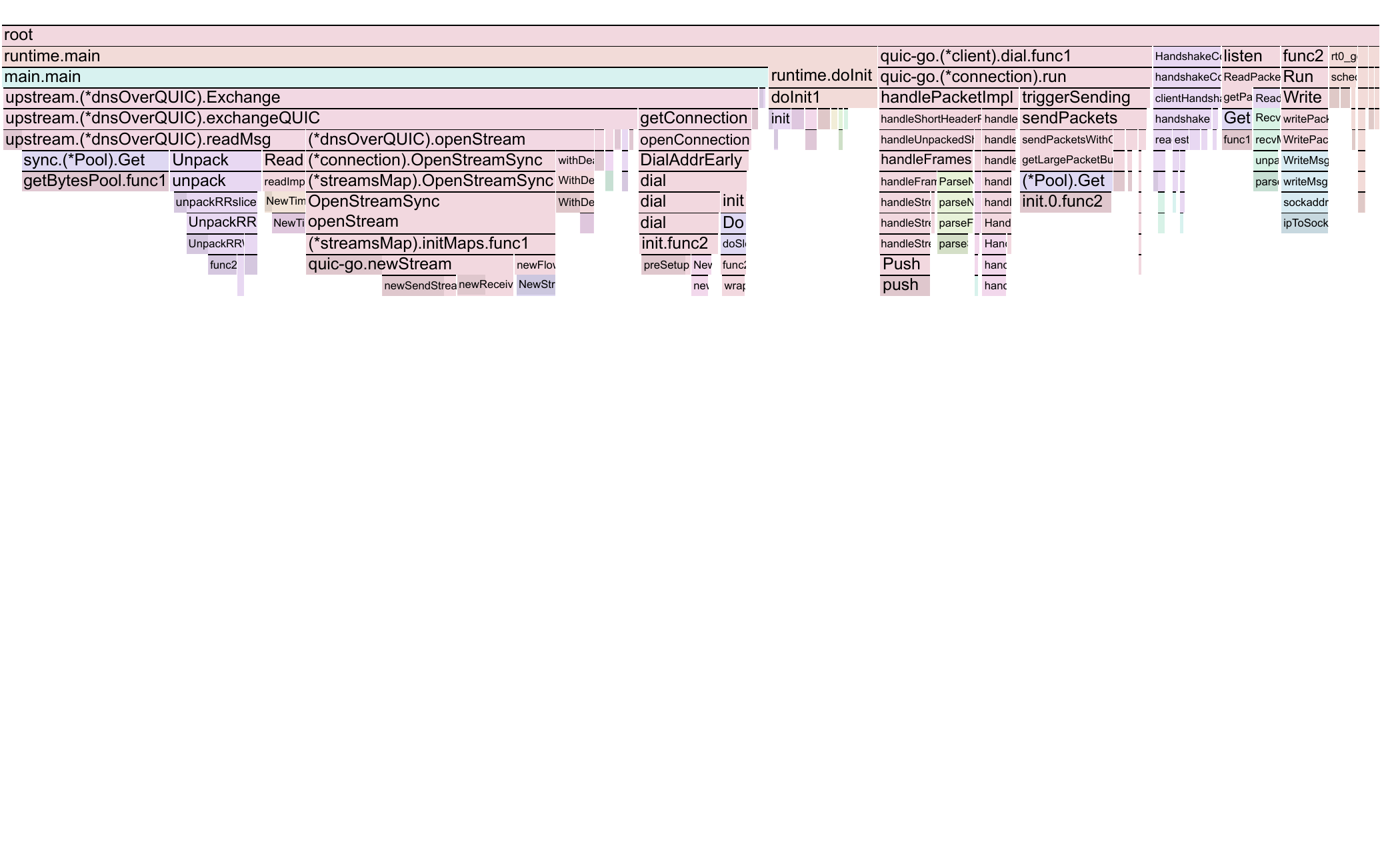}
\vspace{-77mm}
\caption{Stream-Based DoQ Memory Flame Graph (100 Requests)}
\label{flameMem100}
\vspace{-3mm}
\end{figure*}

\begin{figure*}[!b]
\captionsetup[subfloat]{captionskip=-1pt}
\centering
\vspace{-4mm}
\subfloat[DoQ Memory Comparison (100 Requests)]{\label{mem100} \includegraphics[width=0.32\textwidth]{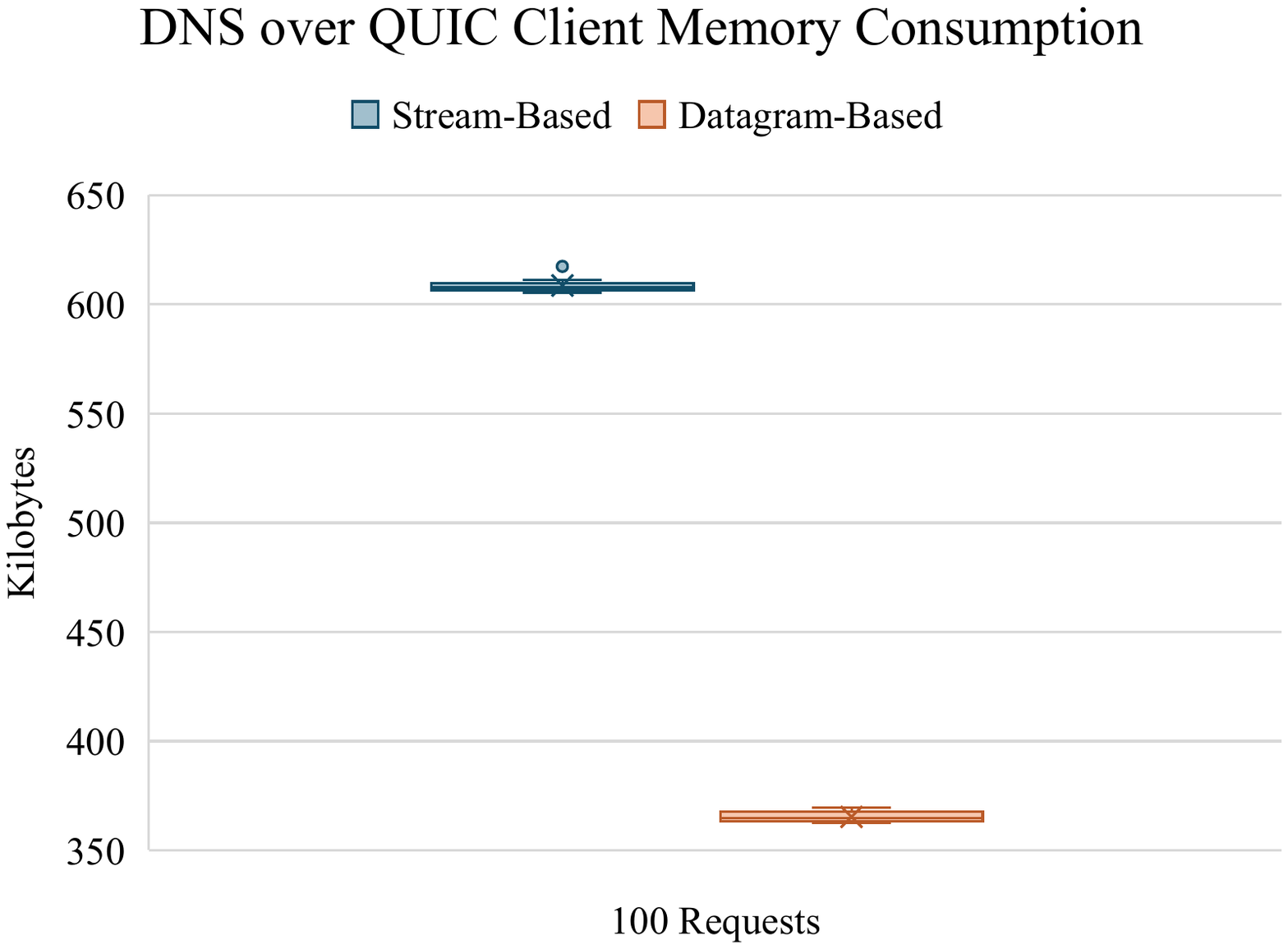}}%
\hfill
\subfloat[DoQ Time Comparison (100 Requests)]{\label{time100} \includegraphics[width=0.32\textwidth]{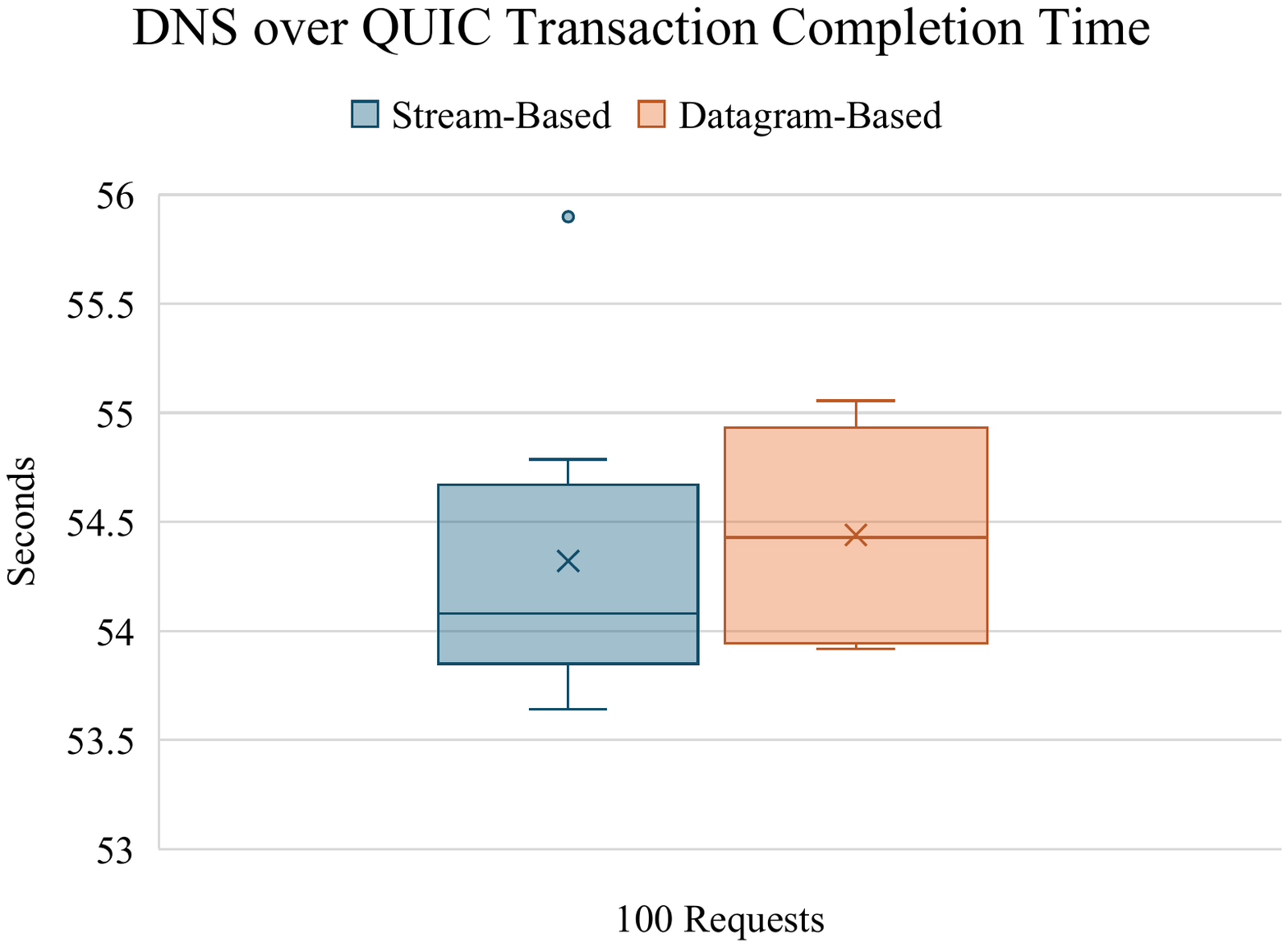}}%
\hfill
\subfloat[DoQ Signaling Overhead (50 Requests)]{\label{byt50} \includegraphics[width=0.32\textwidth]{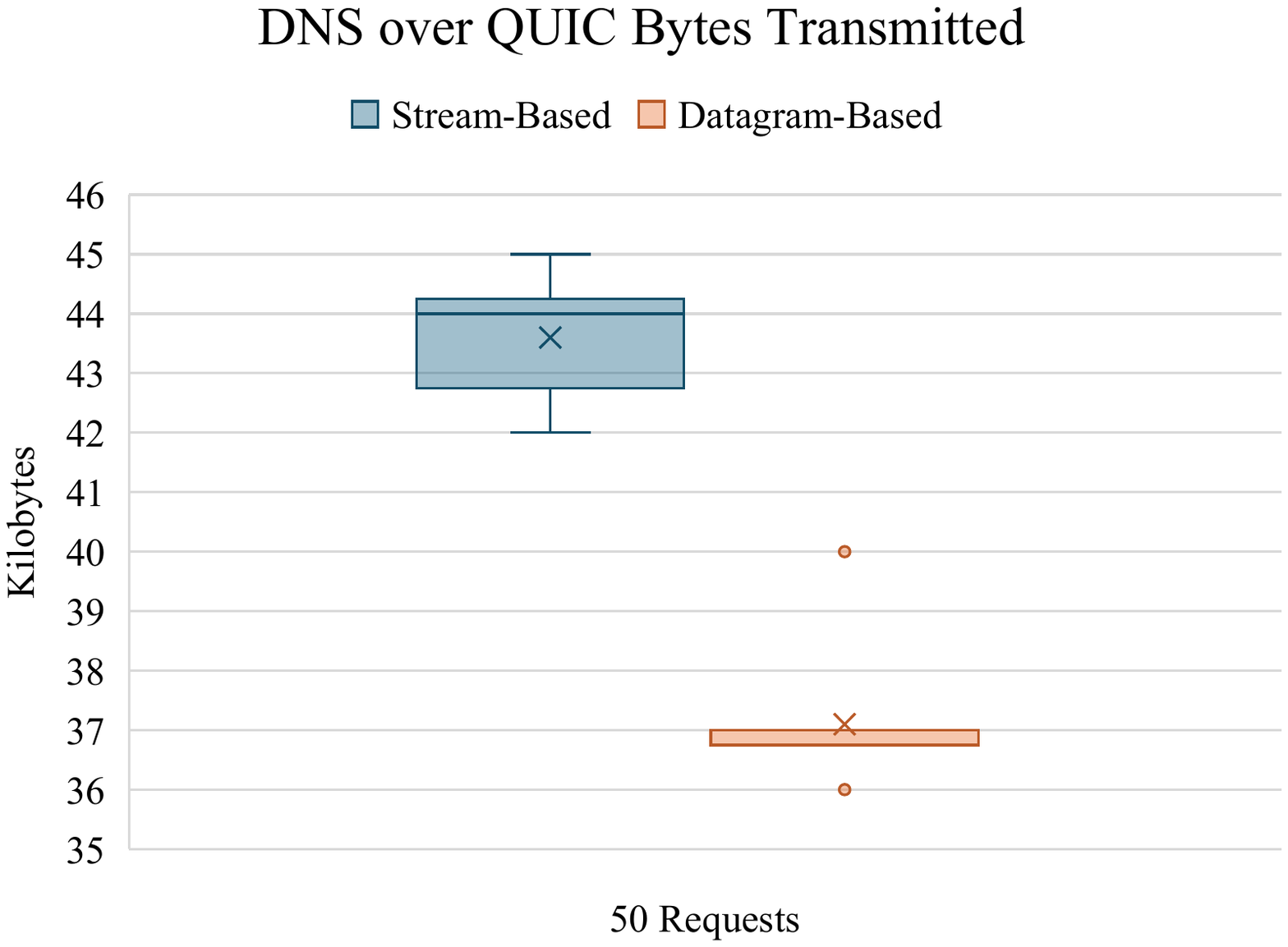}}%
\caption{}
\label{page6Figs}
\end{figure*}

Namely, the \textit{openStream} function which calls QUIC-GO's \textit{openStreamSync} accounted for 63,952 additional bytes of memory allocation, whereas the \textit{getBytesPool} function accounted for 65,552 additional bytes. This explains the delta between the two modes of operation. We note that the \textit{getBytesPool} function was needed to read from QUIC streams. Architecturally, reading from a byte-stream is very different than reading from a standalone datagram. This is reflected in QUIC-GO's APIs for reading from datagrams versus streams -- reading from QUIC streams requires an \textit{io.Reader} and a pre-allocated buffer accommodating the largest message size, whereas \textit{ReceiveDatagram} directly returns the data.

Figure \ref{mem100} depicts a scenario where the number of queries was doubled to demonstrate how memory allocations would change after prolonged usage of the QUIC connection. As the number of queries on the QUIC connection increased, the potential memory savings with datagram-based delivery widened. Datagram and stream delivery modes used 61kB and 152kB more memory, respectively. Connection establishment and handshaking operations were, of course, one-time costs and accounted for a consistent amount of memory under each scenario. The \textit{exchangeQUIC} function was the leading contributor to the discrepancy between each delivery mode. To help illustrate the structure of \textit{dnslookup} and the functions mentioned, a flame graph of stream-based DoQ's memory allocations for 100 queries is provided in Figure \ref{flameMem100}. Each row represents a level of the code's call stack, and the widths of the bars are proportional to each function's memory consumption.


In summary, the creation of QUIC streams for DNS queries had a substantial effect on the client's memory use. This cost was exacerbated by the architectural differences of reading from a byte-stream as opposed to a standalone datagram. Considering the constraints of IoT devices and the frequency of their DNS queries, datagrams appear to be quite advantageous.


\subsection{Total Transaction Completion Time}

The transaction completion time statistics for 100 query-response pairs are provided in Figure \ref{time100}. The boxes of each delivery mode occupy much of the same area, indicating low statistical significance in their respective performance data. Furthermore, the mean value using streams was 54.32 seconds, whereas datagram-based delivery was 54.44 seconds. The median values were 54.08 seconds and 54.42 seconds, respectively. Therefore, we determined that using datagrams in DoQ did not come at any notable performance loss.

\subsection{Client Signaling Overhead}

The bytes signaled between the client and the proxy server for 50 query-response pairs is shown in Figure \ref{byt50}. On average, datagram-based delivery accounted for approximately 6500 fewer bytes. In our experiments, QUIC's Path MTU Discovery \cite{rfc8899} was enabled, which is the default setting. This mechanism sends \texttt{PADDING} frames of increasing size until the path's MTU is fully utilized. For both modes of delivery, these packets accounted for roughly 11.6kB. Connection establishment -- particularly, QUIC version negotiation (Section 6 of RFC 9000 \cite{rfc9000}) -- was another large consumer of network resources which was common to both modes of data delivery. These were one-time costs for each QUIC connection.


\begin{figure*}[t]
\captionsetup[subfloat]{captionskip=-1pt}
\centering
\subfloat[DoQ Signaling Overhead (100 Requests)]{\label{byt100} \includegraphics[width=0.32\textwidth]{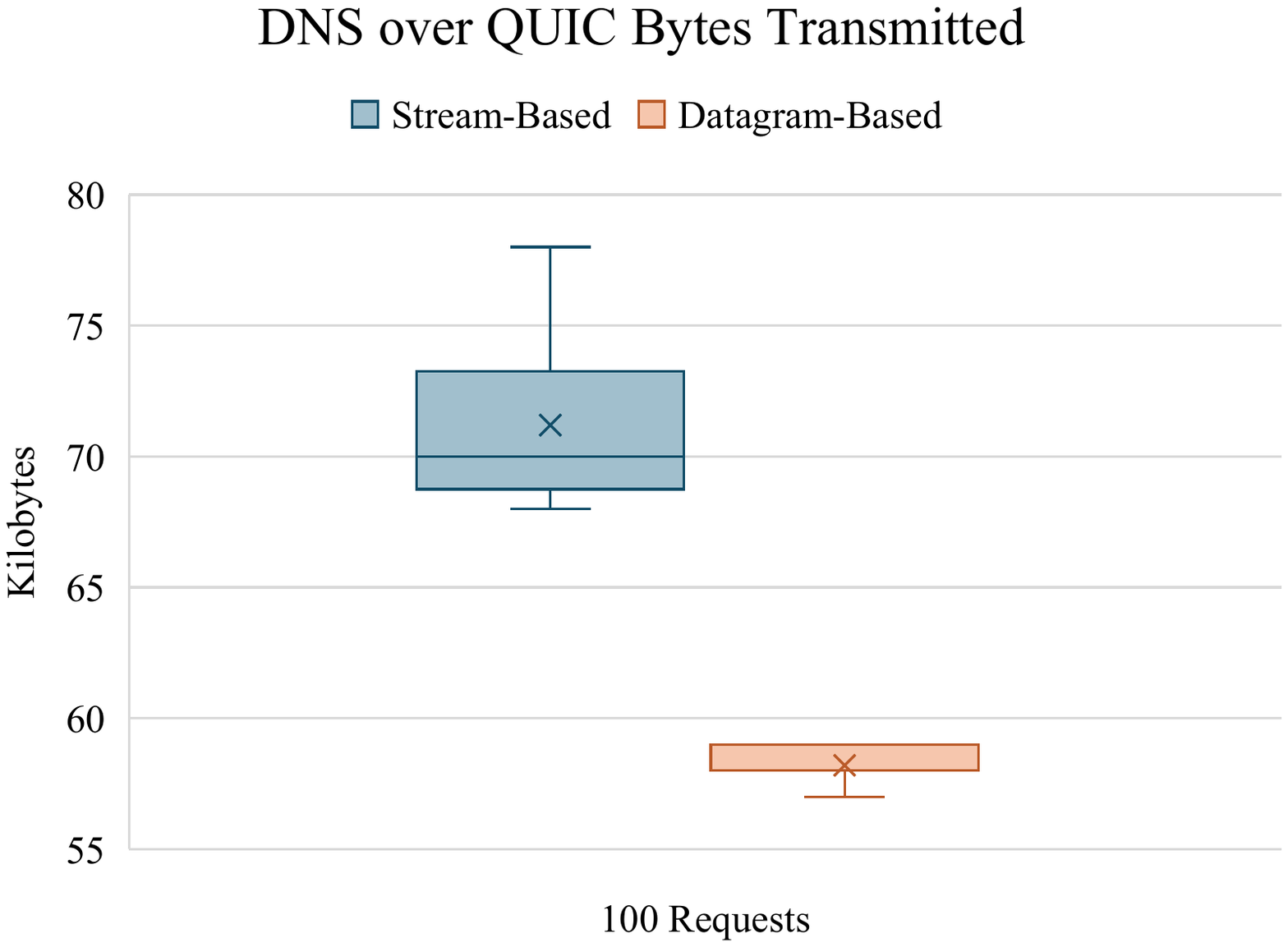}}%
\hfill
\subfloat[DoQ CPU Comparison (50 Requests)]{\label{cpu50} \includegraphics[width=0.32\textwidth]{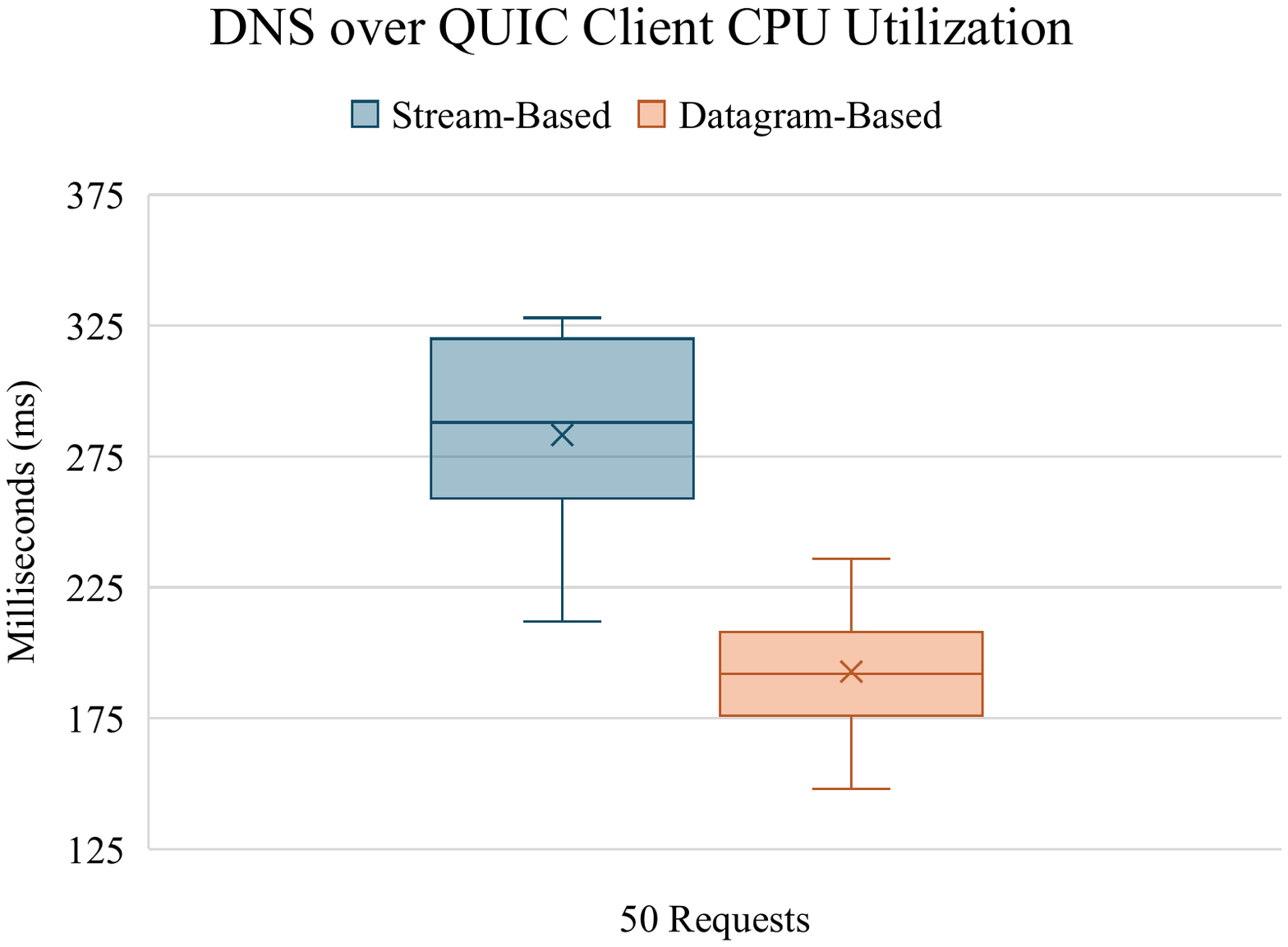}}%
\hfill
\subfloat[DoQ CPU Comparison (100 Requests)]{\label{cpu100} \includegraphics[width=0.32\textwidth]{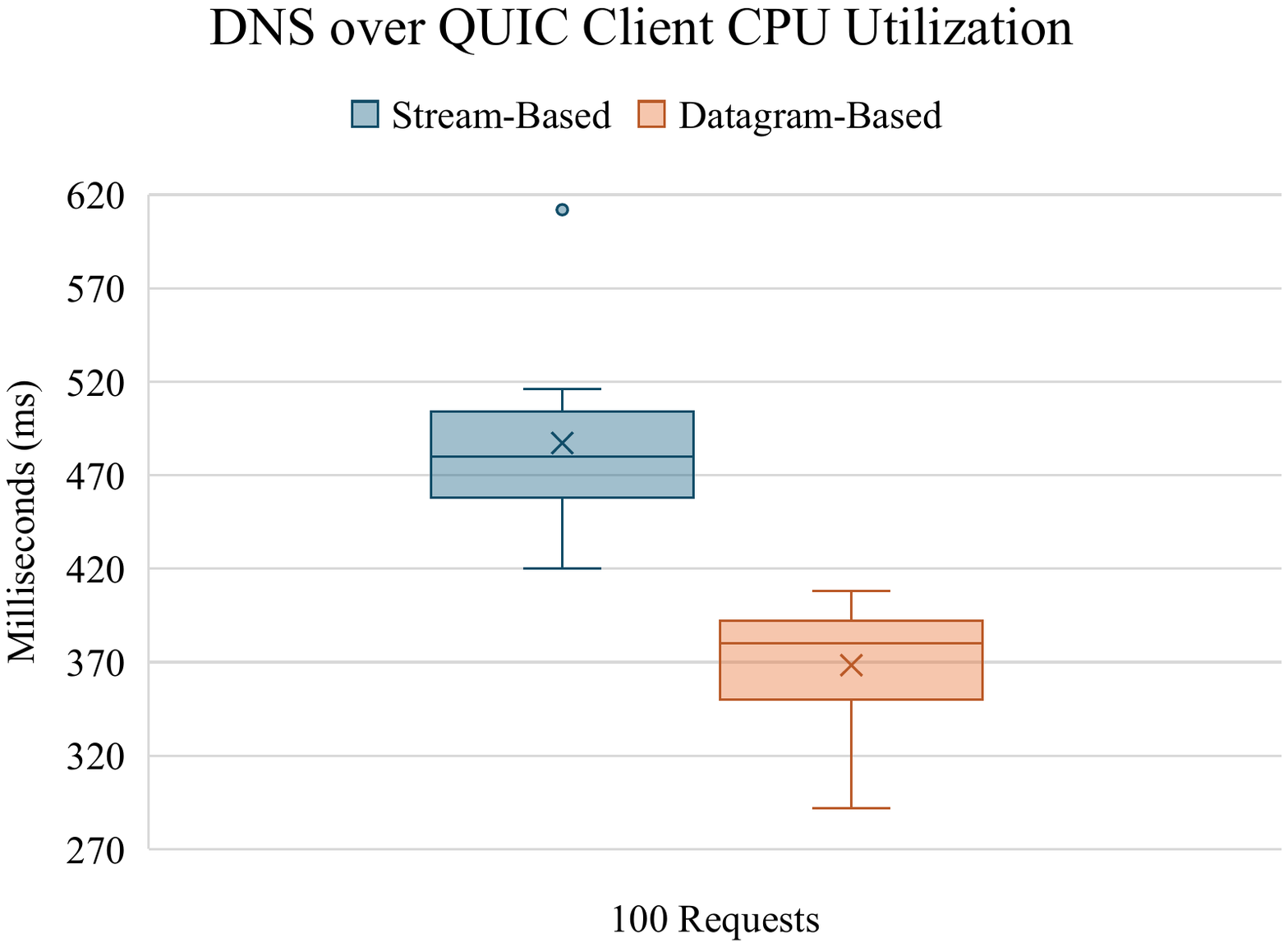}}%
\caption{}
\vspace{-5mm}
\label{page7figs}
\end{figure*}

\texttt{DATAGRAM} frames were very slightly smaller than \texttt{STREAM} frames because they do not require a Stream ID field. Explaining the savings in signaling overhead was the \texttt{MAX\_STREAMS} frames generated in stream-based delivery, as we expected. For every stream which was retired, a \texttt{MAX\_STREAMS} frame was generated by the server. These frames were not coalesced with any other frame types and were sent on their own. Each of these packets accounted for 66 bytes. Worse yet, the \texttt{MAX\_STREAMS} frames were \texttt{ACK} eliciting -- causing even more data exchange from the client as well. In our previous work \cite{saifIoTJourn}, we proposed a two-pronged approach to reduce these effects for IoT applications: i.) by implementing a high-watermark scheme where \texttt{MAX\_STREAMS} are sent when 50\% of the allowance exhausted, and ii.) delaying acknowledgments to cover wider packet ranges (which can be tuned with the \textit{max\_ack\_delay} transport parameter).

When 100 queries were made, which is shown in Figure \ref{byt100}, the mean values of stream and datagram based DoQ differed by 13,000 bytes. This was exactly double the difference presented when 50 query-response pairs were considered and further corroborated that additional signaling is explained by the \texttt{MAX\_STREAMS} behaviors of QUIC streams.


Based on this data, we conclude that QUIC's mechanisms of regulating stream creation -- while sophisticated -- may be excessive for an application like DNS. There is room for some improvement in this area by means of tuning, which can bring the signaling overhead closer to what datagram-based delivery offers. Such tuning options are within the liberties that the QUIC specification gives to implementations. Datagrams had slightly lower overhead and, importantly, did not require additional exchange of further QUIC packets.\vspace{-2mm}

\subsection{Client CPU Usage}

CPU data collected from \textit{pprof} for QUIC connections with 50 query-response pairs is summarized in Figure \ref{cpu50}. The measurements represent each respective mode of operation's time occupying the CPU. Unlike with the memory and signaling measurements, there was overlap in the whiskers between each mode of DoQ delivery. The boxes themselves, however, did not overlap. The mean and median values for each respective mode of operation closely tracked each other.


Stream-based DoQ used the CPU 90.4ms longer, on average. Following the CPU traces, we observed that there was no single consumer which adversely inflated the CPU measurements. Most functions in the traces occupied the CPU for longer in the stream-based delivery mode, and we believe this was simply because more packets were sent, and more memory was allocated over the lifetime of the program.


The CPU data for 100 query-response pairs is shown in Figure \ref{cpu100}. In this case, there was no overlap in either plot's whiskers, suggesting that the delta values between each delivery mode had some dependence on the additional messaging. The overall mean CPU for stream-based delivery increased from 283.2ms to 487.2ms. Stream-based delivery saw an outlier datapoint where 612ms of CPU was used.

Comparing traces between 50 and 100 stream-based delivery exchanges, the \textit{exchangeQUIC} function and its child processes grew from 28ms to 96ms. The QUIC-GO library's \textit{sendPackets} function grew from 20ms to 60ms. The QUIC-GO \textit{listen} function for incoming packets grew from 16ms to 40ms. Lower-level calls to GO's runtime also doubled in usage of the CPU. For datagram-based delivery, on the other hand, the mean CPU value grew from 192.8ms to 368.4ms. Given that double the application data was sent over each QUIC connection (with one-time costs of establishment and path MTU discovery), we deem the increase to be within reason.

\begin{figure}[b]
\centering
\vspace{-4mm}
\includegraphics[width=3.3in]{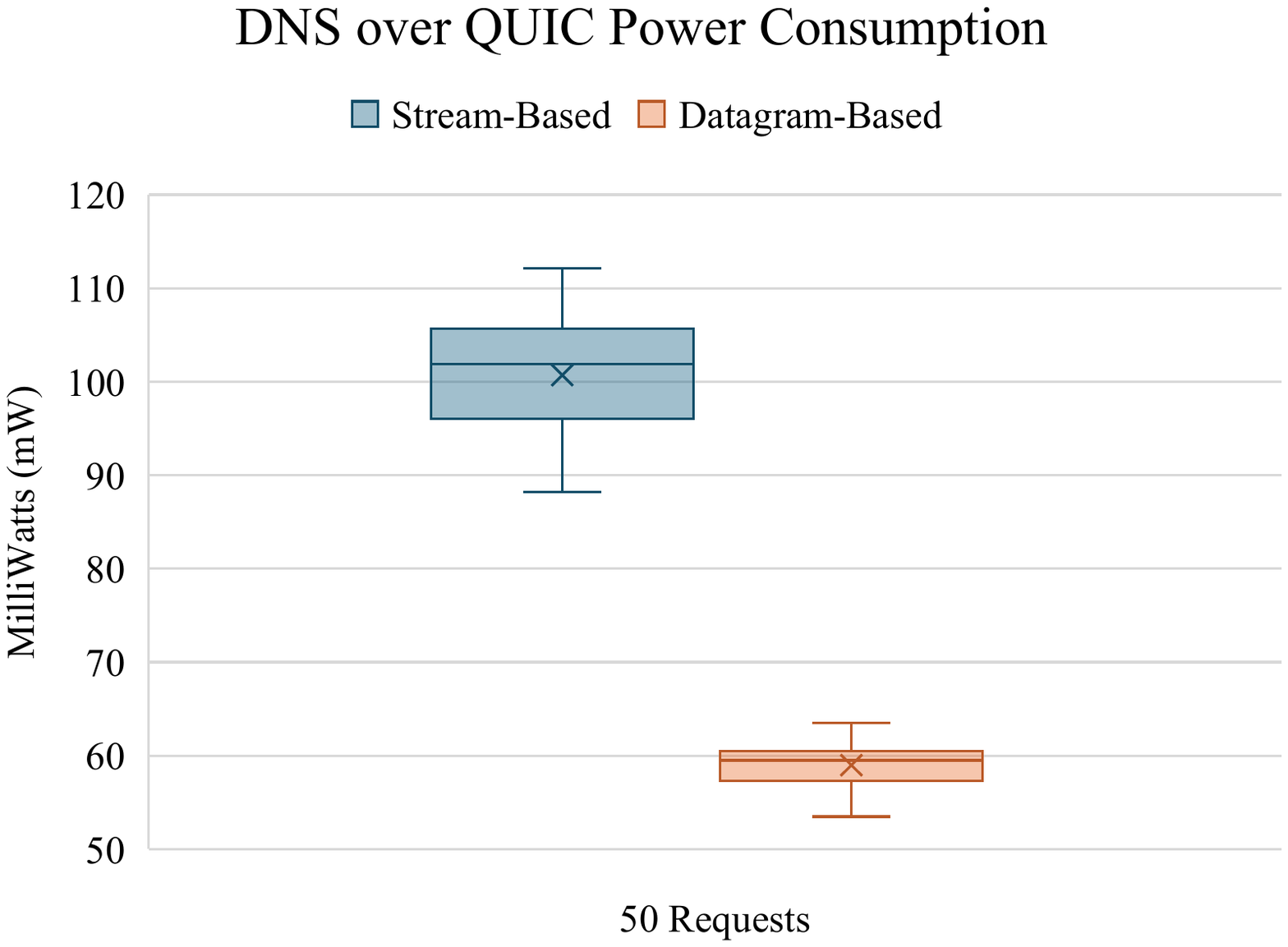}
\vspace{-2mm}
\caption{DoQ Power Consumption Comparison (50 Requests)}
\label{pow50}
\end{figure}

\begin{figure*}[b]
\captionsetup[subfloat]{captionskip=-6pt}
\centering
\vspace{-8mm}
\subfloat[Time Comparison (Latency)]{\label{latTime} \includegraphics[width=0.24\textwidth]{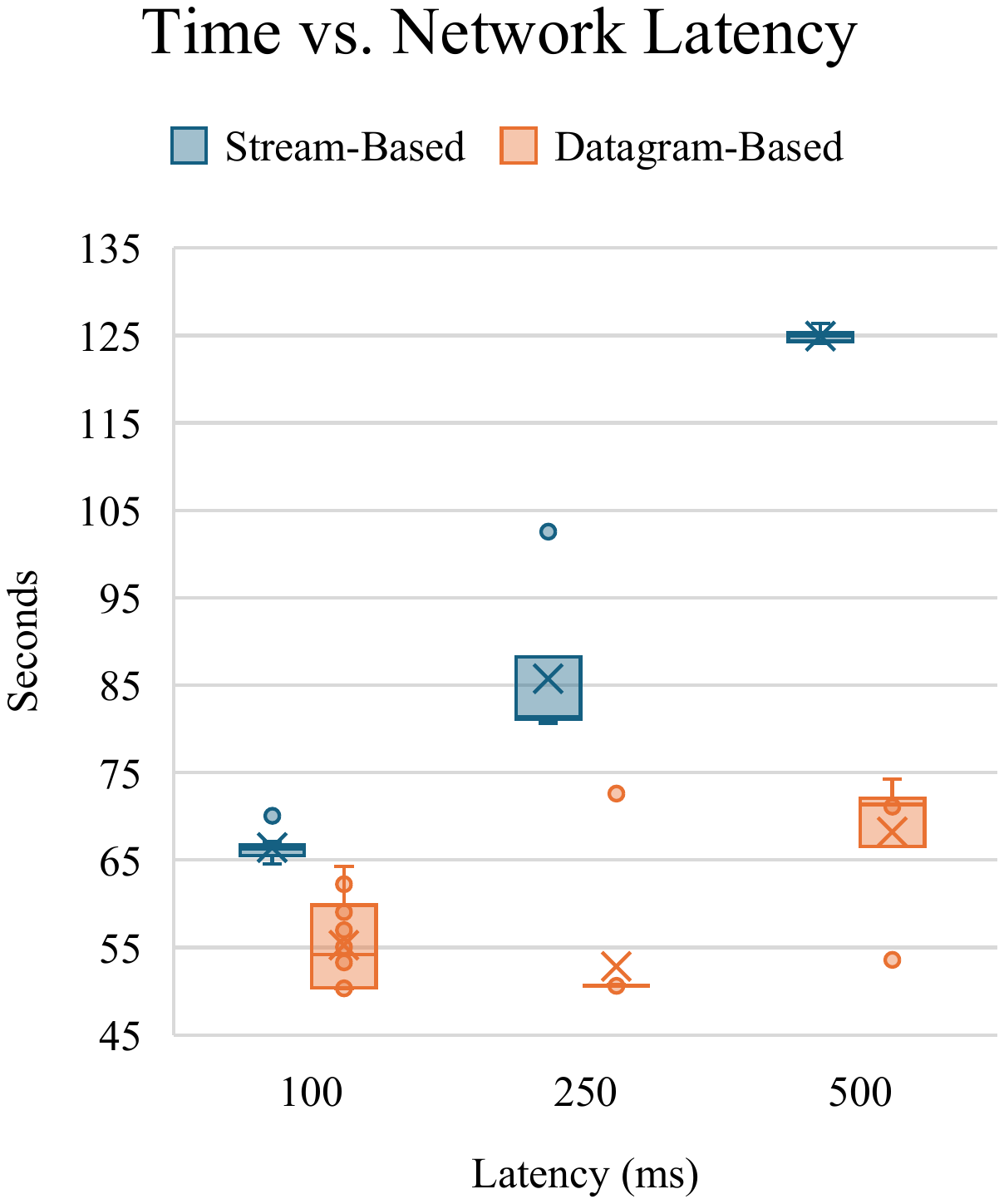}}%
\hfill
\subfloat[Bytes Comparison (Latency)]{\label{latByt} \includegraphics[width=0.24\textwidth]{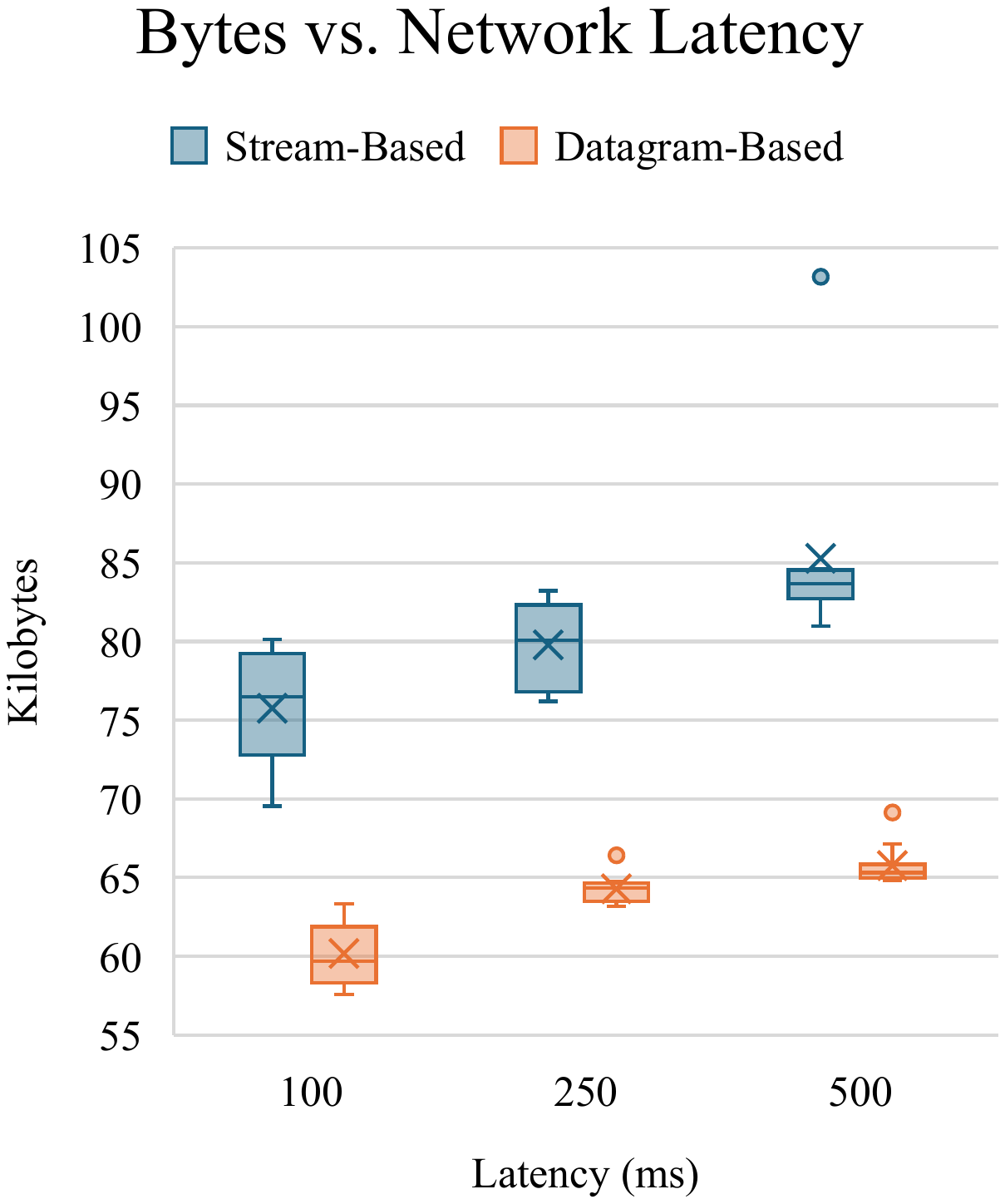}}%
\hfill
\subfloat[Time Comparison (Loss)]{\label{lossTime} \includegraphics[width=0.24\textwidth]{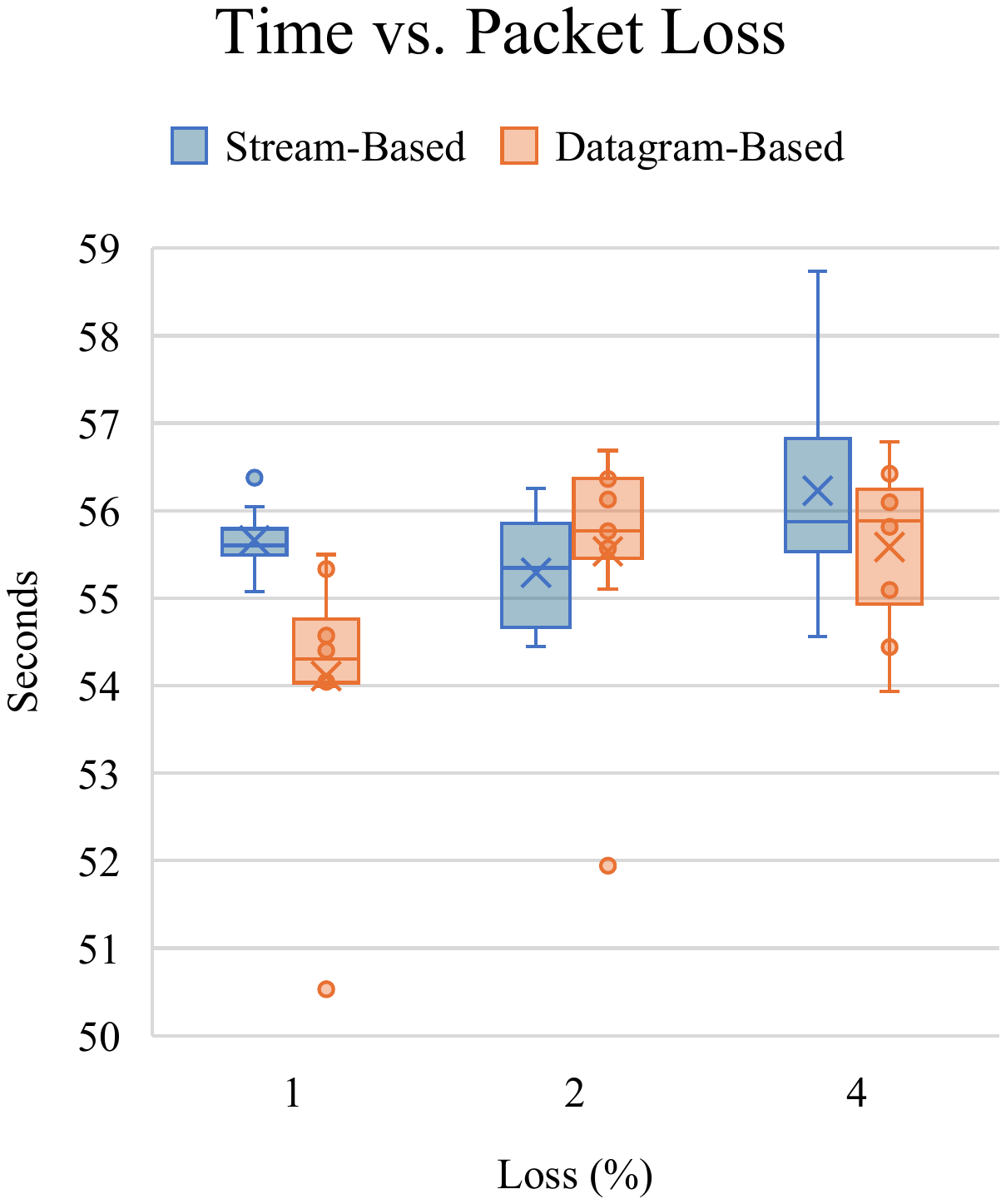}}%
\hfill
\subfloat[Bytes Comparison (Loss)]{\label{lossByt} \includegraphics[width=0.24\textwidth]{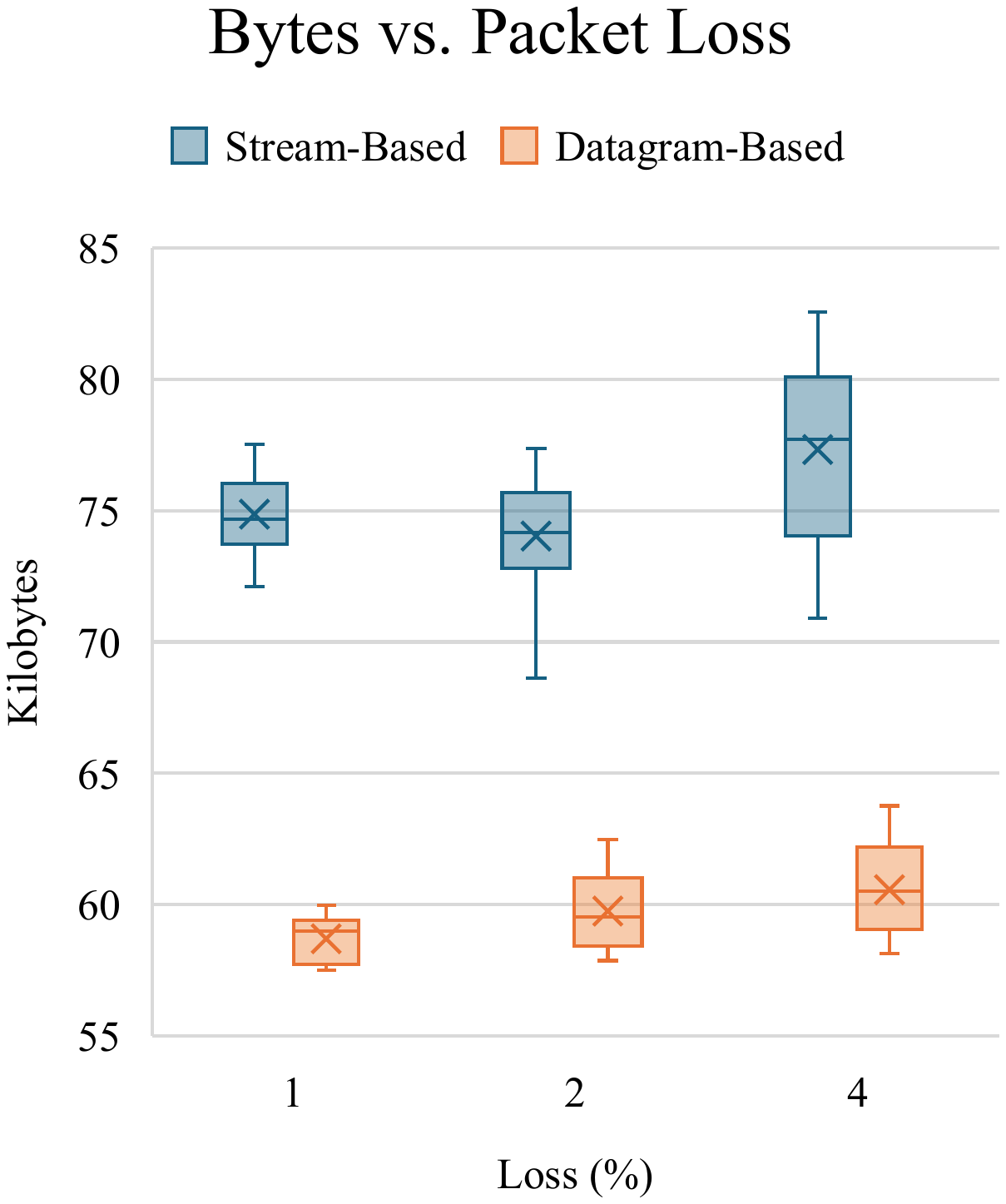}}%
\caption{DoQ NetEm Channel Impairment Comparisons}
\label{latAndLoss}
\end{figure*}

There were two noticeable sources of increase in stream-based DoQ's CPU traces for 100 queries compared to datagram-based: i.) actions associated with opening, closing, and reading from streams accounted for an extra 72ms, and ii.) \textit{mallocgc} (responsible for memory allocation) accounted for an additional 28ms.

This is evidence that managing streams had heavier implications on the client's processor in comparison to datagram-based delivery. Over the course of a connection with more DNS activity, the gap between these modes of delivery became more apparent. Piling on to this, the increased allocation of memory and additional signaling also worked to increase the overall CPU utilization for stream-based delivery.

\subsection{Client Power Consumption}

Linux's \textit{powertop} tool estimates the power draw of all running processes for management and diagnosis purposes. We have leveraged this tool to collect power consumption statistics for each DoQ mode of operation. Every 5 seconds, \textit{powertop} recorded power statistics while DoQ was running. We isolated all processes related to our DoQ program and calculated the average power draw per experimental iteration. The results for 50 queries are shown in Figure \ref{pow50}.

Datagram-based delivery proved to have meaningfully less power consumption. The mean value across all 10 iterations was 59 mW whereas stream-based delivery's mean was recorded as 100.7 mW. Stream-based delivery appeared to exhibit more variability as well, indicated by the lengths of its box and whiskers. Data was also collected for 100 queries, but the results were nearly identical to those shown in Figure \ref{pow50}. Because queries are sent at the same rate, average power consumption is not affected. Therefore, we have not provided a power plot for 100 queries. We note that because the program ran twice as long for 100 queries, however, the energy expenditure would be double.

\subsection{Latency and Packet Loss}
 
We also assessed the resilience of each operation mode under unpredictable network channel conditions. There were two reasons motivating this investigation: i.) such conditions prevalent in IoT \cite{rfc7228}, and ii.) datagram-based DoQ involves non-confirmed data transmission (like standard DNS over UDP), whereas stream-based DoQ is confirmed. We used NetEm to increase the network latency -- with a uniform distribution -- and introduced random packet loss as well. A fixed number of 100 queries was tested in these configurations.

Figures \ref{latTime} and \ref{latByt} show the transaction execution time and signaling overhead for each DoQ transmission mode, respectively. The RTT between the client and \textit{dnsproxy} server was increased by 100ms, 250ms, and 500ms. The time results of stream-based DoQ were more dramatically affected by increasing RTT, which is understandable. Each stream-based transaction consists of more round-trips: requiring a client \texttt{ACK} of the DNS response, \texttt{MAX\_STREAMS} frames from the server, and another client \texttt{ACK}. Our proposed datagram-based DoQ required only a single round-trip. 

The bytes transmitted for datagram-based DoQ increase with latency because the backoff timer was triggered more frequently. Stream-based DoQ also saw an increase in bytes transmitted -- this was because of Probe Timeouts (PTOs). QUIC sends PTOs when \texttt{ACK}-eliciting packets are not acknowledged within an expected period of time so that a connection may recover from the loss of tail packets or \texttt{ACK}s \cite{rfc9002}. This time period is derived from the smoothed RTT and the RTT variance, along with other factors. QUIC relies on an initial RTT guess when no previous samples are known, and so (even with a uniform latency distribution), it can take time for the smoothed RTT to converge. In the packet captures we collected, we found several PTOs, consisting of \texttt{PING} and \texttt{PADDING} frames averaging 1400 bytes each.

Execution time and signaling overhead data under increasing packet loss are shown in Figures \ref{lossTime} and \ref{lossByt}, respectively. As packet loss grew, both modes of DoQ experienced higher variability and thus wider boxes and whiskers. Because stream-based DoQ sends more packets than datagram-based, it had higher exposure to packet loss, and this is reflected in both figures. The median execution times of datagram-based DoQ grew at a higher rate than stream-based, because of its more rudimentary retry mechanism. However, both plots retained significant box and whisker overlap. The signaling overhead of each DoQ operation mode increased roughly by the same margin. These experiments show that datagram-based DoQ has advantages even under unpredictable network conditions.

\section{Conclusion}

In this paper, we have evaluated DoQ under scenarios where multiple queries were sent in a QUIC connection -- which to our knowledge has not been performed before. We hypothesized that, due to DNS's high message volume, frequent creation of QUIC streams would result in a high CPU, memory, and signaling footprint for IoT devices. Reductions in such areas are paramount to help make DoQ more feasible in IoT, while conserving bandwidth and device energy.

To address this challenge, we designed and implemented an alternative mode of DNS delivery using QUIC's unreliable datagram extension. Our design is non-disruptive to the original DoQ specification, as clients can choose whether or not to opt in to datagram-based delivery. While the benefits are most pronounced on IoT devices, datagram-based delivery can be advantageous to other DNS clients as well. One limitation of datagram-based DoQ is that a client cannot cancel a transaction like it would be able to with \texttt{STOP\_SENDING} frames pertaining to streams.

This work is important because DNS is a critical Internet service where low-latency results have been favored over providing security and privacy. This makes DNS an easy target for eavesdropping, which has proven to be effective towards pinpointing what kinds of IoT devices exist on a network. DNS over QUIC transport offers modern features which address these privacy concerns while maintaining lower latency than predecessor technologies like DoT and DoH.

Our results collected from a Raspberry Pi Zero client show a decisive savings in memory allocations when comparing datagram-based DoQ to stream-based. In our setup, dealing with QUIC streams accounted for roughly 2.5kB more memory allocations per DNS transaction than datagram-based DoQ. Because of QUIC's need to signal \texttt{MAX\_STREAMS} frames for stream regulation, it was also found that additional packets were transmitted, amounting to approximately 130 more bytes per DNS query. CPU utilization and power consumption of each delivery method also reflected these findings. These effects quickly accumulated into a compelling advantage for datagram-based DoQ as more DNS transactions transpired. It is also important to note that the resource savings we have outlined are largely independent of the inter-dispersal time of DNS queries, but rather accumulate with each transaction.

Given these reductions and the fact that there was no performance degradation in transaction completion time (even with high RTT and packet loss), we conclude that datagram-based delivery can indeed facilitate DoQ's feasibility for IoT and it is worth consideration as an extension to RFC 9250.

\bibliographystyle{ieeetr}
\bibliography{references.bib}

\begin{IEEEbiography}[{\includegraphics[width=1in,height=1.25in,clip,keepaspectratio]{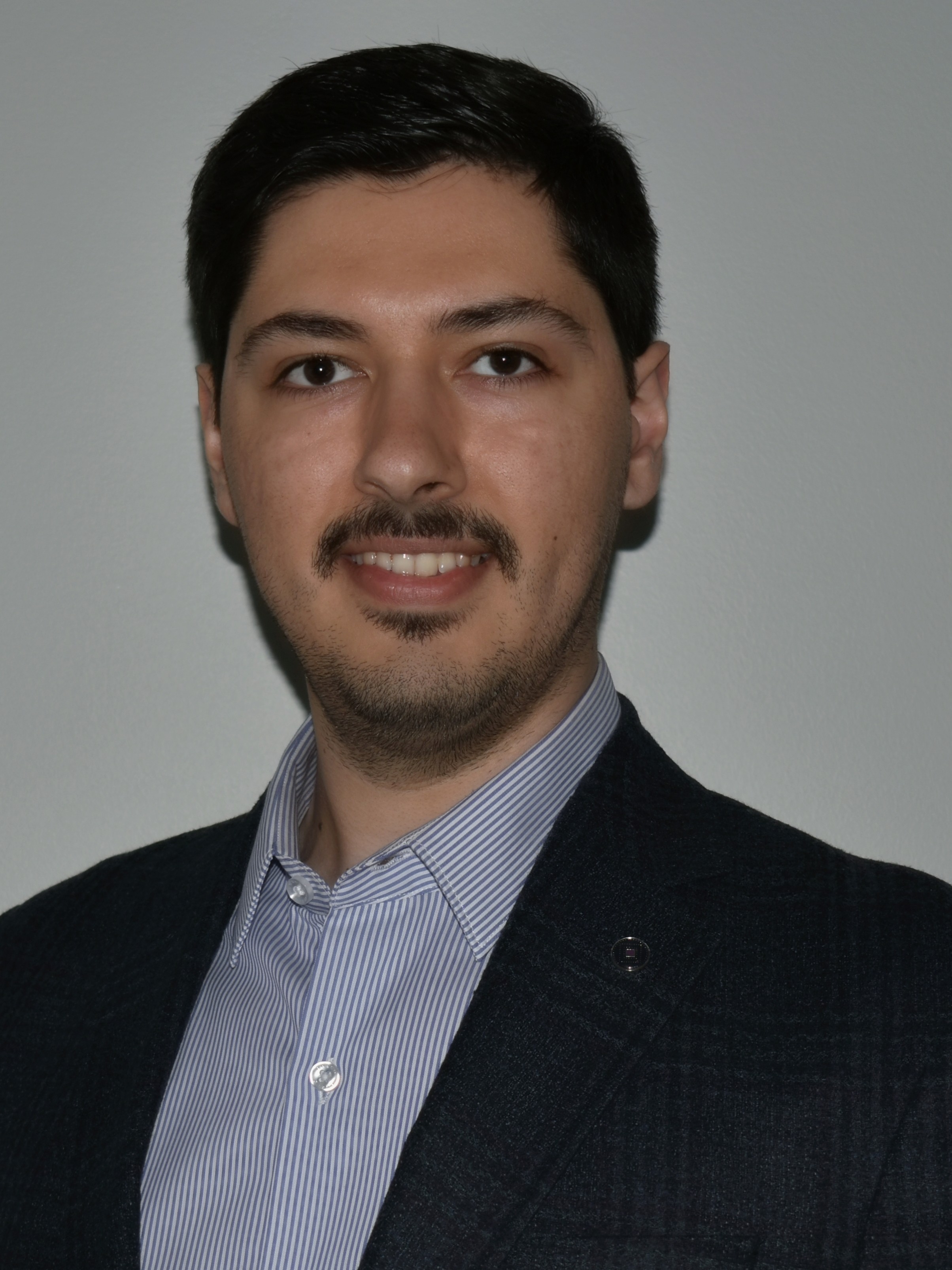}}]{Darius Saif}
is a Ph.D. candidate and member of the Next Generation Networking Lab at Carleton University in Ottawa, Canada. Since 2016, he has been a part of Nokia's Network Infrastructure R\&D Department, working on software automation and verification for Deep Packet Inspection (DPI) capabilities in residential, mobile, and fixed wireless networks. Darius completed his Bachelor's of Applied Science degree with distinction in the field of Electrical \& Computer Engineering at the University of Windsor in 2016.
\end{IEEEbiography}

\begin{IEEEbiography}[{\includegraphics[width=1in,height=1.25in,clip,keepaspectratio]{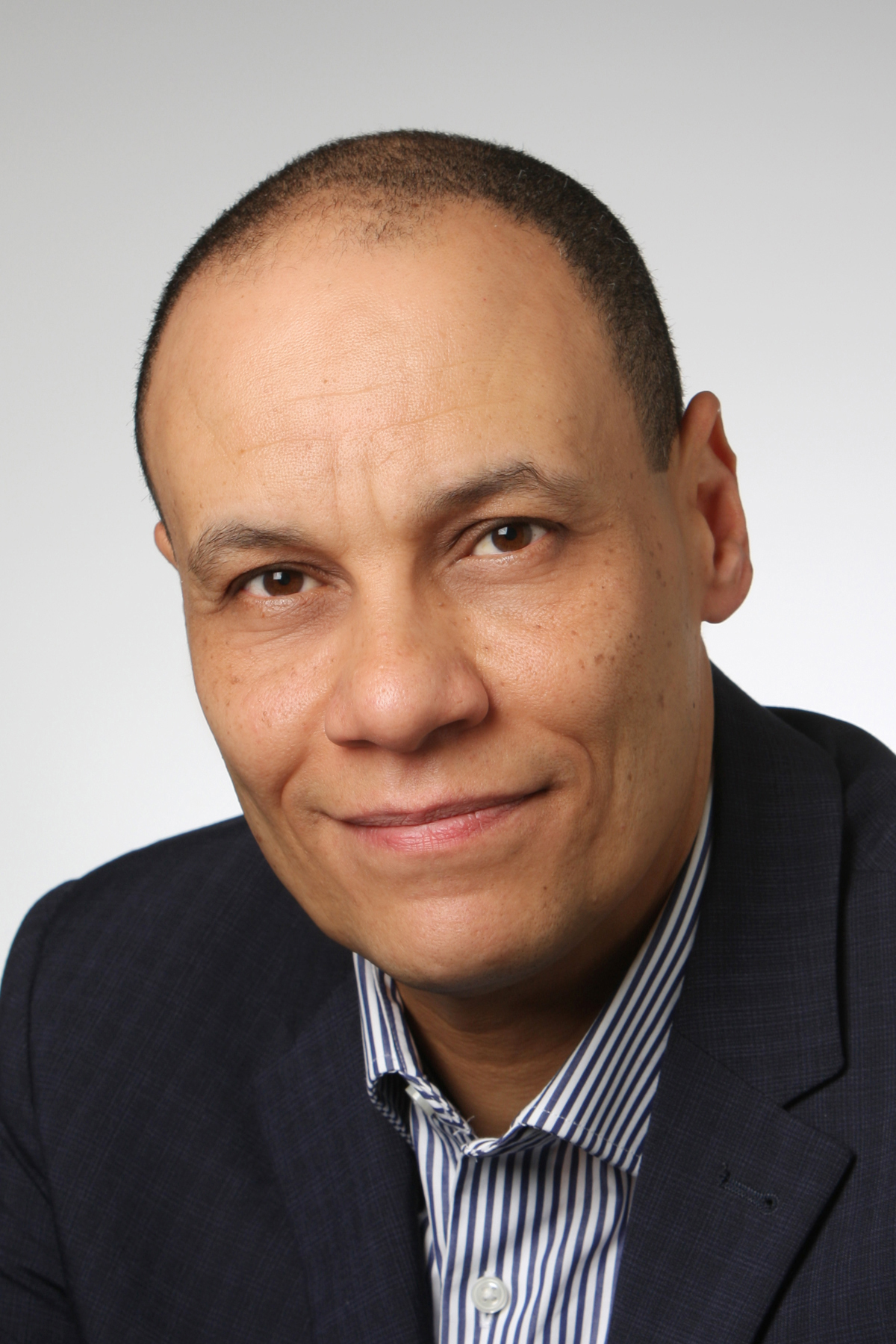}}]{Dr. Ashraf Matrawy}
is a Professor at Carleton University. He is a senior
member of the IEEE and a licensed professional engineering in Ontario.
His area of interest is reliable and secure computer networking. He is
the winner of multiple best paper awards, the faculty graduate mentoring
award in 2019, the Faculty of Engineering and the University Research
Awards in 2021 and 2022, and the IEEE Ottawa Section Outstanding
Engineering Educator Award in 2021. His research has been supported by
government and industrial partners such as Alcatel-Lucent, TELUS, and
NSERC.
\end{IEEEbiography}

\vfill
\end{document}